\documentclass[a4paper,11pt]{article}
\pdfoutput=1 

\usepackage{jcappub} 

\usepackage[T1]{fontenc} 

\title{\boldmath Traversable Wormhole in Einstein 3-Form Theory With Self-Interacting Potential}

\author[a,b]{Mariam Bouhmadi-L\'opez }
\author[c]{Che-Yu Chen}
\author[d,e,1]{Xiao Yan Chew \note{Corresponding author.}}
\author[f,g]{Yen Chin Ong}
\author[d,e]{Dong-han Yeom}

\affiliation[a]{Department of Physics, University of the Basque Country UPV/EHU, Bilbao 48080, Spain}
\affiliation[b]{IKERBASQUE, Basque Foundation for Science, Bilbao 48011, Spain}
\affiliation[c]{Institute of Physics, Academia Sinica, Taipei 11529, Taiwan}
\affiliation[d]{Department of Physics Education, Pusan National University, Busan 46241, Republic of Korea}
\affiliation[e]{Research Center for Dielectric and Advanced Matter Physics, Pusan National University, Busan 46241, Republic of Korea}
\affiliation[f]{Center for Gravitation and Cosmology, College of Physical Science and Technology, Yangzhou University, \\180 Siwangting Road, Yangzhou City, Jiangsu Province  225002, China}
\affiliation[g]{Shanghai Frontier Science Center for Gravitational Wave Detection, School of Aeronautics and Astronautics, Shanghai Jiao Tong University, Shanghai 200240, China}

\emailAdd{mariam.bouhmadi@ehu.eus}
\emailAdd{b97202056@gmail.com}
\emailAdd{xychew998@gmail.com}
\emailAdd{ycong@yzu.edu.cn}
\emailAdd{innocent.yeom@gmail.com}

\abstract{
We numerically construct a symmetric wormhole solution in pure Einstein gravity supported by a massive $3$-form field with a potential that contains a quartic self-interaction term. The wormhole spacetimes have only a single throat and they are everywhere regular and asymptotically flat. Furthermore, their mass and throat circumference increase almost linearly as the coefficient of the quartic self-interaction term $\Lambda$ increases. The amount of violation of the null energy condition (NEC) is proportional to the magnitude of $3$-form, thus the NEC is less violated as $\Lambda$ increases, since the magnitude of $3$-form decreases with $\Lambda$. In addition, we investigate the geodesics of particles moving around the wormhole. The unstable photon orbit is located at the throat. We also find that the wormhole can cast a shadow whose apparent size is smaller than that cast by the Schwarzschild black hole, but reduces to it when $\Lambda$ acquires a large value. The behavior of the innermost stable circular orbit around this wormhole is also discussed. The results of this paper hint toward the possibility that the 3-form wormholes could be potential black hole mimickers, as long as $\Lambda$ is sufficiently large, precisely when NEC is weakly violated.
}

\begin{document}
\maketitle
\flushbottom

\section{Introduction}

Wormholes are well-known peculiar solutions in general relativity (GR) and many modified/alternative theories of gravity. They typically consist of two ``mouths'' connected through a ``bridge''. The section of the bridge with minimal surface area is known as the ``throat''. If a wormhole allows an observer to enter one mouth and exit another, it is said to be ``traversable''. This can happen only when the bridge is timelike.
The ongoing quest for wormhole solutions is of course partly motivated by humanity's hope for interstellar travel, as traversable wormholes in principle provide a shortcut connecting far-flung regions of the universe. More realistically, however, we hope that investigating wormholes can push our understanding of gravity and spacetime structures (its possible geometries and topologies) to its extreme.

The classic example of a non-traversable wormhole is the Einstein-Rosen bridge which can be obtained from a coordinate transformation of the Schwarzschild black hole \cite{Einstein:1935tc}, i.e. it \emph{is} part of the structures in a maximally extended Kruskal-Szekeres manifold. It is no wonder then any observer who attempts to cross such a wormhole would be terminated at the Schwarzschild singularity. 
As is well-known, in GR, the construction of traversable wormholes typically requires the violation of energy conditions, in order to prevent the collapse of the throat \cite{Visser:1995cc,visser1,visser2,MTY}. See \cite{1405.0403} for more discussions on energy conditions.

The classic and simplest example of traversable wormholes that violates the energy condition is the Ellis wormhole whose throat is supported by a phantom field \cite{Ellis:1973yv,Ellis:1979bh,Bronnikov:1973fh}, which is a scalar field with an opposite sign for the kinetic term. Such matter fields with negative energy or negative mass density are typically problematic in one way or another. Negative energy is often associated with a runaway instability (hence the importance of the positive energy theorem in classic GR). For example, we see the danger of the phantom field in the case of the Big Rip singularity in cosmology \cite{0302506}.

One can also construct solutions of traversable wormholes in modified gravity in which any phantom-like matter field is no longer needed \cite{Harko:2013yb}. For example, this has been done in Einstein-Gauss-Bonnet-scalar theory \cite{Kanti:2011yv,Antoniou:2019awm,Brihaye:2020dgo,Ibadov:2020btp}, Lovelock gravity \cite{Chakraborty:2021qwl,Harko:2013yb}, Horndeski gravity \cite{Mironov:2018uou,Volkova:2019kyd,Korolev:2020ohi}, 
 vector-tensor theories with Abelian gauge symmetry breaking \cite{Li:2020jyf} and hybrid metric-Palatini gravity \cite{Rosa:2018jwp,Rosa:2021yym}. In some modified theories of gravity, traversable wormholes can even form via quantum tunnelings and bubble materializations \cite{Battarra:2014naa,Bouhmadi-Lopez:2018sto,Tumurtushaa:2018agq,Chew:2020lkj}. Recently, we also note that a traversable wormhole in GR can be constructed in the Einstein-Dirac \cite{Blazquez-Salcedo:2019uqq} and Einstein-Maxwell-Dirac theories \cite{Blazquez-Salcedo:2020czn,Konoplya:2021hsm}. Indeed, the latter approach has the merit of Occam's razor: GR is a well-tested theory, so it would be interesting to find traversable wormhole solutions in GR with ``physically reasonable'' matter field to support the throat. Such a matter field should preferably possess a correct sign for its kinetic energy though necessarily still violate some energy conditions. It would be even better if such a matter field is in some sense natural (e.g., it can also be applied to explain cosmological puzzles such as the accelerated expansion of the Universe).

One natural candidate is the 3-form field which is ubiquitous to string theory \cite{Copeland:1994km,Lukas:1996iq,Ovrut:1997ur,Farakos:2017ocw} within a cosmological framework and beyond. This is not surprising as a 3-form behaves like a cosmological constant in absence of a potential \cite{Duff:1980qv}. Already back in the late nineties, 3-form fields were proposed as a
mechanism to induce inflation within a quantum cosmological setup \cite{Turok:1998he}.  In a much more updated scenario and using current observational data, 3-forms have been deeply analysed as candidates to drive the primordial early inflationary era and have proven to successfully fit the data \cite{Germani:2009iq,Koivisto:2009sd,Germani:2009gg,Koivisto:2009fb,Koivisto:2009ew,DeFelice:2012jt,DeFelice:2012wy,Kumar:2014oka,Mulryne:2012ax,SravanKumar:2016biw}. It does not come as a big surprise that 3-forms can as well describe the late-time speed up \cite{Koivisto:2009fb,Koivisto:2009ew}. What it certainly  comes as a surprise is that 3-forms can behave as phantom-like matters with a correct kinetic term  \cite{Morais:2016bev,Bouhmadi-Lopez:2016dzw}. In some cases 3-forms can induce dark energy singularities that can be cured at the quantum level \cite{Bouhmadi-Lopez:2018lly}. This repulsive effect can cure black hole singularities as we showed recently in \cite{Bouhmadi-Lopez:2020wve} and support repulsive points at infinity \cite{Bouhmadi-Lopez:2016dzw}. For a recent review about the state of the art on cosmology, we refer the reader to \cite{CANTATA:2021ktz}. It seems, therefore, natural to look for wormhole solutions supported by 3-forms as we do on the current work. 

In this work, in line with this line of thinking, we construct traversable wormhole solutions in pure GR supported by a $3$-form field which violates the null energy condition (NEC). Note that wormhole solutions in the presence of a 3-form field were already previously found in \cite{Barros:2018lca}. Nevertheless, an anisotropic fluid was required to support the wormhole in addition to the 3-form field. Our approach does not require an additional fluid or scalar field and as such is ``minimal''. We remark that 3-form black hole solutions \cite{Barros:2020ghz,Bouhmadi-Lopez:2020wve} and 3-form stellar models \cite{Barros:2021jbt} have also been discussed.

Our paper is organized as follows. In Sec.~\ref{sec.the}, we introduce our theoretical setup which includes a $3$-form field as the matter content in GR, the metric ansatz for the wormhole spacetime, and its corresponding geometry. The wormhole solution is derived numerically and is presented in Sec.~\ref{sec.result}. The violation of the NEC will be discussed. Furthermore, we discuss the geodesic equations of particles moving around the wormhole in Sec.~\ref{sec.geo}. The locations of the photon sphere, the size of the wormhole shadow, and the radius of innermost stable circular orbit (ISCO) will be investigated. Finally, we conclude our work and discuss the outlook in Sec.~\ref{sec:con}. We will work in the units $G=c=1$.

 \section{Theoretical Framework}\label{sec.the}

\subsection{The Three Form}

First, let us consider Einstein gravity minimally coupled with a 3-form potential $A_{\alpha \beta \gamma}$ \cite{Barros:2018lca}. The total action reads
\begin{equation} \label{EHaction}
 S=  \int d^4 x \sqrt{-g}  \left(  \frac{R}{16 \pi} + \mathcal{L}_{A}  \right)  \,,
\end{equation}
where $R$ is the Ricci scalar. The Lagrangian for $A_{\alpha \beta \gamma}$ is denoted by $\mathcal{L}_{A}$. Explicitly, it is
\begin{equation}
  \mathcal{L}_{\text{A}} =- \frac{1}{48} F^2  - V(A^2) \,, 
\end{equation}
with the square denoting the contraction of all indices, i.e., $A^2=A_{\alpha \beta \gamma} A^{\alpha \beta \gamma}$. In the 3-form Lagrangian above, $V \equiv  V(A^2)$ is the potential function. The 4-form $F=dA$ is the field strength tensor, whose components can be written as
\begin{equation}
F_{\alpha \beta \gamma \delta} = \nabla_{\alpha} A_{\beta \gamma \delta} -  \nabla_{\delta} A_{\alpha \beta \gamma} +  \nabla_{\gamma} A_{\delta \alpha \beta} -  \nabla_{\beta} A_{\gamma \delta \alpha} \,.
\end{equation}
 
The Einstein equation can be obtained by varying the action $S$ with respect to the metric $g_{\mu \nu}$,
\begin{equation} \label{einstein_eqn}
 R_{\mu \nu} - \frac{1}{2} g_{\mu \nu} R =  \kappa  T_{\mu \nu}  \,,
\end{equation}
where $\kappa=8 \pi$ and $T_{\mu \nu}$ is the stress-energy tensor for $A_{\alpha \beta \gamma}$ given by
\begin{equation}
 T_{\mu \nu} = \frac{1}{6} F_{\mu \alpha \beta \gamma} F_{\nu}\,^{\alpha \beta \gamma} + 6 \frac{\partial V}{\partial A^2} A_{\mu \alpha \beta} A_{\nu}\,^{\alpha \beta} + g_{\mu \nu} \mathcal{L}_{A} \,.\label{TMUNU}
\end{equation}
On the other hand, the equation of motion for the 3-form field is obtained by varying $S$ with respect to $A_{\alpha \beta \gamma}$. This yields 
\begin{equation}  \label{eom_3form}
 \nabla_{\mu} F^{\mu \alpha \beta \gamma} = 12 \frac{\partial V}{\partial A^2} A^{\alpha \beta \gamma} \,.
\end{equation}

Generally speaking, the 3-form field can be defined as $A_{\alpha \beta \gamma} = \sqrt{-g} \epsilon_{\alpha \beta \gamma \delta} B^{\delta}$
where, in a static and spherically symmetric spacetime with a radial coordinate $r$, the 1-form $B^{\delta}$ can be parametrized by a single function $\chi(r)$ as
\begin{equation}
 B^\delta = \left( 0, \chi(r) , 0, 0  \right)  \,.
\end{equation}
We shall adopt the following rather simple potential to construct the wormhole solution,
\begin{equation}
 V(A^2) = \mu^2 A^2 + \lambda A^4\,,\label{Va2}
\end{equation}
where $\mu$ effectively stands for the mass of the 3-form, and $\lambda$ corresponds to the coefficient of the quartic self-interaction term. Note also that the self-interaction term is defined by $A^4 := (A_{\alpha \beta \gamma} A^{\alpha \beta \gamma})^2$.

\subsection{The Metric Ansatz}

In this work, we will focus on the construction of a static and spherically symmetric wormhole spacetime. More explicitly, we follow the approach of \cite{Dzhunushaliev:2017syc} by employing the following metric ansatz, 
\begin{equation}
  ds^2 = - P(r)^2 dt^2 + d r^2 + R(r)^2 (d \theta^2 + \sin^2 \theta d \varphi^2)  \,,\label{metricansatz}
\end{equation}
where $P(r)$ is a dimensionless redshift function and $R(r)$ is the shape radial function, which stands for the circumferential radius of the wormhole. We require that the wormhole spacetime possesses two asymptotically flat regions as $r \rightarrow \pm \infty$, i.e., $P(r)^2\rightarrow1$ and $R(r)^2\rightarrow r^2$ for $r \rightarrow \pm \infty$. In order to have a globally regular wormhole solution, the function $R(r)$ should not possess any zero. Instead it should possess at least one local minimum, which would correspond to the wormhole throat. If $R(r)$ possesses a local maximum, that maximum would be referred to as the equator, at which the wormhole geometry acquires the largest surface area (except at sufficiently large distances near spatial infinities). For a wormhole with such a geometry, the equator is normally sandwiched between two throats. 

From the metric ansatz \eqref{metricansatz}, one can calculate the non-zero components of the 3-form field $A_{\alpha\beta\gamma}$ as follows: 
\begin{equation}
A_{t\theta\varphi}=A_{\varphi t\theta}=A_{\theta\varphi t}=-A_{t\varphi\theta}=-A_{\theta t\varphi}=-A_{\varphi\theta t}=\sqrt{-g}\chi(r)\,.
\end{equation}
Using these components, we obtain $A^2=-6\chi^2$ and $F^2=-24Y^2$, where
\begin{equation}
Y:=\chi'+\chi\left(\frac{P'}{P}+\frac{2R'}{R}\right)\,.
\end{equation}
In the previous expression, the prime denotes the derivative with respect to $r$. Note that we also have the following identity
\begin{equation}
\frac{\partial V}{\partial A^2}=-\frac{1}{12}\frac{1}{\chi}\frac{\partial V}{\partial\chi}\,.
\end{equation}
Furthermore, one can define the energy density and the pressures of the 3-form using the expression of the stress-energy momentum tensor \eqref{TMUNU} as follows \cite{Barros:2018lca}:
\begin{align}
\rho_A:=-T^t\,_t&=\frac{Y^2}{2}-\chi\frac{\partial V}{\partial\chi}+V\,,\label{rhoA}\\
p_{1A}:=T^r\,_r&=-\frac{Y^2}{2}-V\,,\qquad p_{2A}:= T^\theta\,_\theta=T^\varphi\,_\varphi=-\rho_A\,.\label{pA}
\end{align}
In 3-form cosmology, the dynamics of the 3-form field can also be described by a scalar degree of freedom, which is similar to the role played by the $\chi$ field here. However, in that case, the scalar degree of freedom describes the evolution of the field in time, and it parametrizes the $t$-component of the 1-form $B^\delta$. For the metric ansatz \eqref{metricansatz} under consideration here, the spacetime is static and the $\chi$ field describes the behavior of the field in space. This leads to the fact that the expressions for the energy density and the pressures of the 3-form in terms of $\chi$ are different from those in 3-form cosmology expressed in terms of the corresponding scalar degree of freedom. See for example Ref.~\cite{Morais:2016bev}.

By substituting the metric \eqref{metricansatz} into Einstein equations and the 3-form field equation, we obtain the following set of ordinary differential equations (ODEs),
\begin{align}
 R'' + \frac{R'^2}{2 R} - \frac{1}{2 R} = - \kappa \frac{R}{4} \chi'^2 &- \kappa \left( R' + \frac{R P' }{2 P}    \right) \chi \chi'   \nonumber \\
& \quad - \kappa \left[ \frac{R'^2}{R} + \frac{P' R'}{P} + \frac{R P'^2}{4 P^2} + 3 \mu^2 R \right] \chi^2  + 54 \kappa \lambda R \chi^4 \,, \label{anw1} \\
\frac{2 P' R'}{P R} - \frac{1}{R^2} + \frac{R'^2}{R^2} = - \frac{\kappa}{2} \chi'^2  & - \kappa \left( \frac{2 R'}{R} + \frac{P'}{P}     \right) \chi \chi'  \nonumber \\
& \quad - \kappa \left[ \frac{2 R'^2}{R^2} + \frac{2 P' R'}{ P R} + \frac{P'^2}{2 P^2} - 6 \mu^2  \right] \chi^2   -36 \kappa \lambda \chi^4  \,,  \label{anw2} \\
P'' + \frac{P}{R} R'' + \frac{P' R'}{R} =  - \kappa \frac{P}{2} \chi'^2  &- \kappa \left( \frac{2 P R'}{R} + A'     \right) \chi \chi' \nonumber \\
& \quad - \kappa \left[ \frac{2 P R'^2}{R^2} + \frac{2 P' R'}{R} + \frac{P'^2}{2 P} + 6 \mu^2 P  \right] \chi^2+  108 \kappa \lambda P \chi^4 \,,  \label{anw3} \\
\chi'' + \left( \frac{2 R'}{R} + \frac{P'}{P}  \right) \chi' - & \left( \frac{2 R'^2}{R^2} + \frac{P'^2}{P^2} - \frac{P''}{P} - \frac{2 R''}{R}   \right) \chi = 12 \mu^2 \chi - 144 \lambda \chi^3  \,.  \label{anw4}
\end{align}
Eqs.~\eqref{anw1}, \eqref{anw2}, and \eqref{anw3} are derived from the $tt$, $rr$, $\theta\theta$ components of the Einstein equations, respectively. On the other hand, Eq.~\eqref{anw4} is obtained from the $3$-form equation \eqref{eom_3form}. 

For the sake of later convenience, we introduce the following dimensionless variables to re-scale the ODEs above \eqref{anw1}-\eqref{anw4}:
\begin{equation}
 r = \frac{x}{\mu} \,,  \quad R= \frac{X}{\mu} \,,  \quad \lambda = \kappa \mu^2 \Lambda \,, \quad \chi = \frac{\Psi}{ \sqrt{\kappa} }\,.
\end{equation}
After some algebraic manipulations, Eqs. \eqref{anw1}, \eqref{anw3} and \eqref{anw4} lead to the following final set of ODEs, which we will integrate numerically to obtain the wormhole solution:
\begin{align}
 P'' &=  \frac{P \left( X'^2-1 \right)}{X^2} - 6 P \Psi^2 + 72 P \Lambda \Psi^4  \,, \\
X'' &=  \frac{P' X'}{ P} - 6  X \Psi^2 + 72 \Lambda X \Psi^4 \,, \label{eq20}\\
\Psi'' &=-  \left( \frac{P'}{P} + \frac{2 X'}{X} \right) \Psi'  + \left(  \frac{1}{X^2}+\frac{X'^2}{X^2} + \frac{P'^2}{P^2} - \frac{2 P' X'}{P X}   \right) \Psi + 12  \Psi - 144 \Lambda \Psi^3  + 18 \Psi^3 - 216 \Lambda \Psi^5 \,.
\end{align}
Note that Eq. \eqref{anw2} is a constraint and will be used to monitor the precision of the numerics later. Also, hereinafter the prime denotes the derivative with respect to the dimensionless variable $x$.

For simplicity, we integrate the ODEs from $x=0$ to infinity by assuming that the wormhole spacetime is symmetric with respect to the radial coordinate $x=0$, thus $x=0$ can be either a throat or an equator. Based on this assumption, we require that the first-order derivatives of all functions vanish at $x=0$, such that all functions acquire their extrema at that point, namely, 
\begin{equation}
P'(0)=X'(0)=\Psi'(0)=0\,.\label{iniderivative}
\end{equation} 
The symmetry of the spacetime with respect to $x=0$ enables us to construct a wormhole spacetime with two identical asymptotically flat regions. As just mentioned, the condition Eq.~\eqref{iniderivative} implies that the wormhole spacetime has a throat or an equator at $x=0$. On the other hand, to ensure the asymptotic flatness of the solution at the spatial infinity $(x \rightarrow \pm \infty)$, the functions should behave as $P \rightarrow 1$, $X \rightarrow |x|$ and $\Psi \rightarrow 0$ at the asymptotic regions. 

Technically, to proceed with the numerical integrations, we have to impose boundary conditions at $x=0$, then integrate outward to get the expressions of the metric functions and the 3-form field, as functions of $x$. The boundary conditions require essentially the determination of the values of these functions at $x=0$. In order to study the behaviour of the solutions at $x=0$, we make the following series expansions near $x=0$,
\begin{equation}
 P = p_0 +  p_2 x^2 + ...... \,, \quad X = X_0 + X_2 x^2 + ...... \,,\quad \Psi = \Psi_0  + \Psi_2 x^2 + ......\,.
\end{equation}
Furthermore, the series expansions can determine the relation between the parameters $p_0$, $X_0$, $\Psi_0$ and $\Lambda$ because not all of them are free parameters. Hence, by substituting the series expansions into the constraint equation \eqref{anw2}, one gets the following relation,
\begin{equation}
X_0 = \frac{1}{\Psi_0 \sqrt{-6 + 36 \Psi^2_0 \Lambda}} \,,\label{X0eqvsL}
\end{equation}  
where $X_0$ is interpreted as the circumferential radius at $x=0$. The condition that the wormhole possesses a throat at $x=0$ is guaranteed by the flaring-out condition (see Eq.~\eqref{eq20})
\begin{equation}
 X_2 =  3 X_0 \Psi_0^2 \left( 12 \Lambda \Psi^2_0 -1   \right)  > 0 \,.
\end{equation}
If $X_2 < 0$, this means the wormhole possesses an equator at $x=0$ and, for a viable traversable wormhole, there must be another throat located somewhere else in $0 < x < \infty$. This scenario is possible if one considers a more complicated 3-form potential $V(A^2)$. For simplicity. we shall not consider this possibility in this paper and will only focus on the potential \eqref{Va2}. Now, it is very clear that the only free variable is $\Lambda$ because the values of $p_0$ and $\Psi_0$ can be determined from the asymptotic flatness condition. 

The set of ODEs is solved numerically by the shooting method with standard Runge-Kutta Fourth order method. Before ending this section and exhibiting our main results, we will define the mass of the wormhole. This is essential when we discuss some physical observables later, such as the shadow cast by the wormhole and the ISCO in the spacetime. Since the spacetime is asymptotically flat, as will be shown by our numerical results in the next section, we find it convenient to define the mass of the wormhole via the Misner-Sharp definition \cite{Misner:1964je}, 
\begin{align}
 M &= \frac{R_0}{2} - \frac{\kappa}{2}  \int^{\infty}_{R_0} {T^t}_t R^2 dR \,, \nonumber\\ \nonumber 
\Rightarrow M \mu &=  \frac{X_0}{2} -  \frac{1}{2} \int^{\infty}_{0}  \left[ - \frac{1}{2} X^2 \Psi'^2  - \left( 2 X' + X \frac{P'}{P}   \right) X \Psi \Psi' + 108 \Lambda X^2 \Psi^4  \right. \nonumber  \\
&  \qquad  \qquad \qquad \qquad  \qquad \left. - \frac{1}{2} \left( 4 X'^2 + X^2 \frac{P'^2}{P^2} + \frac{4 X P' X'}{P} + 12 X^2     \right) \Psi^2    \right]   X' dx  \,,\label{whmass}
\end{align}
where $M\mu$ is dimensionless.

\begin{figure}[t!]
\centering
\mbox{ 
(a)  \includegraphics[angle =-90,scale=0.3]{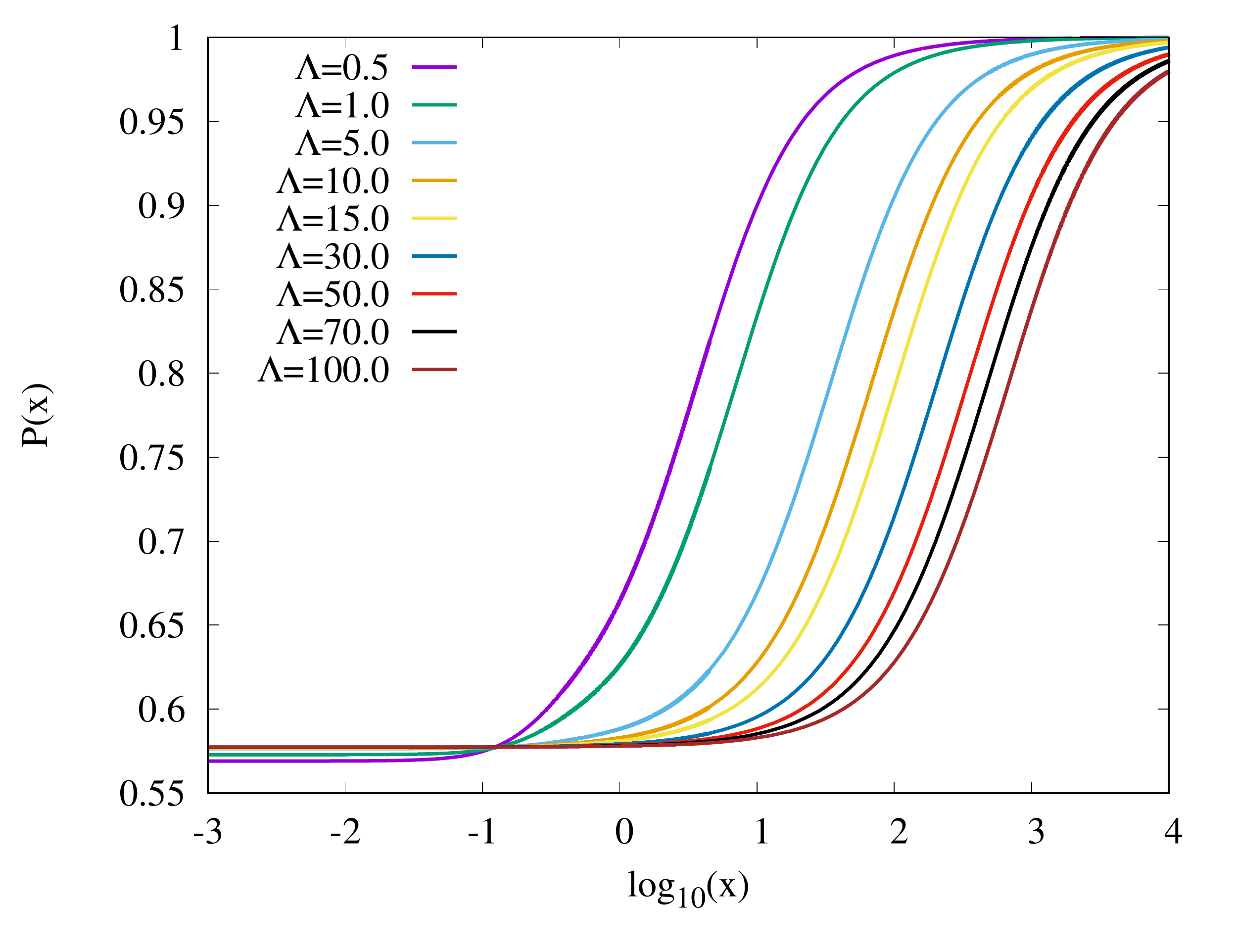}
(b) \includegraphics[angle =-90,scale=0.3]{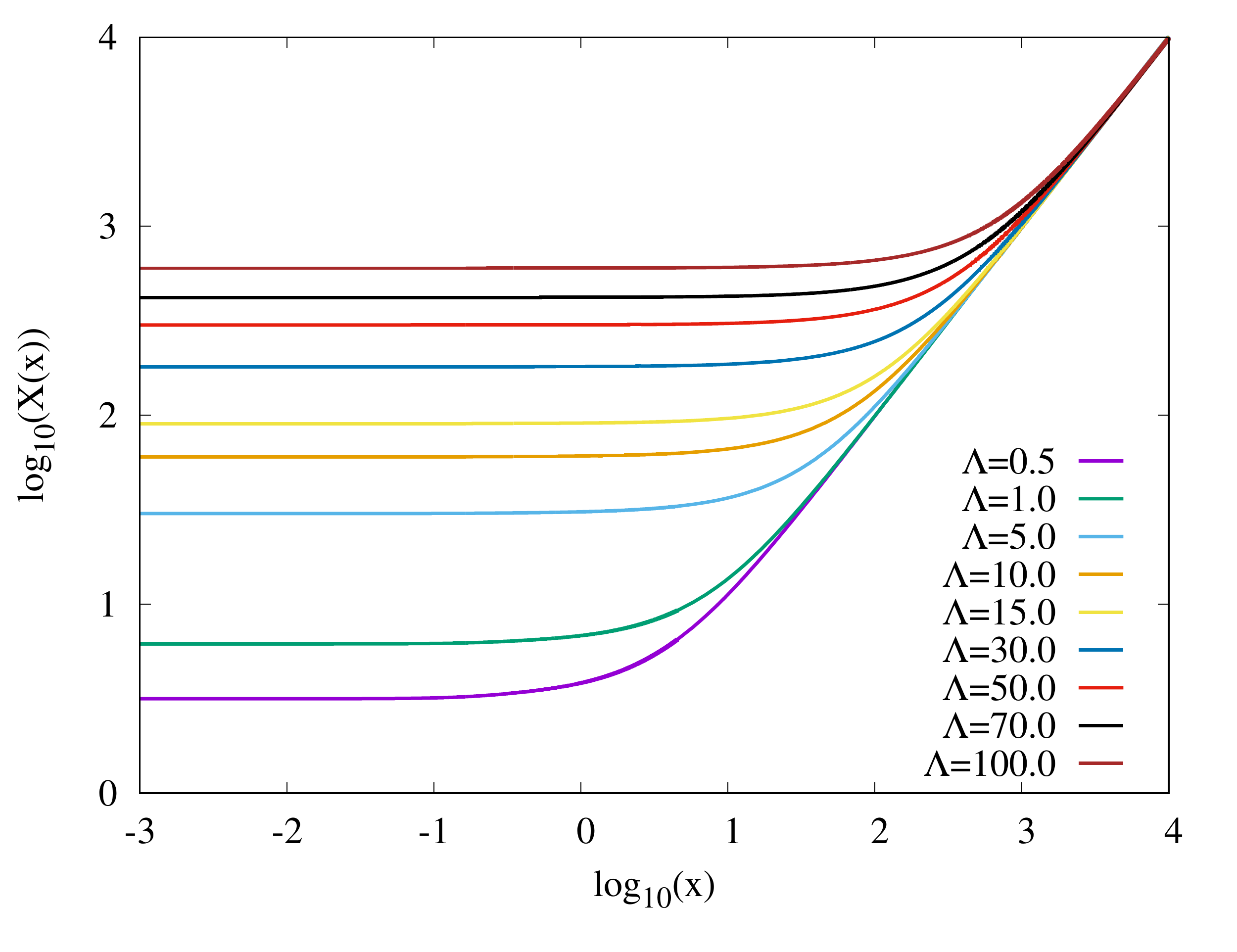} 
}
\mbox{
(c)   \includegraphics[angle =-90,scale=0.3]{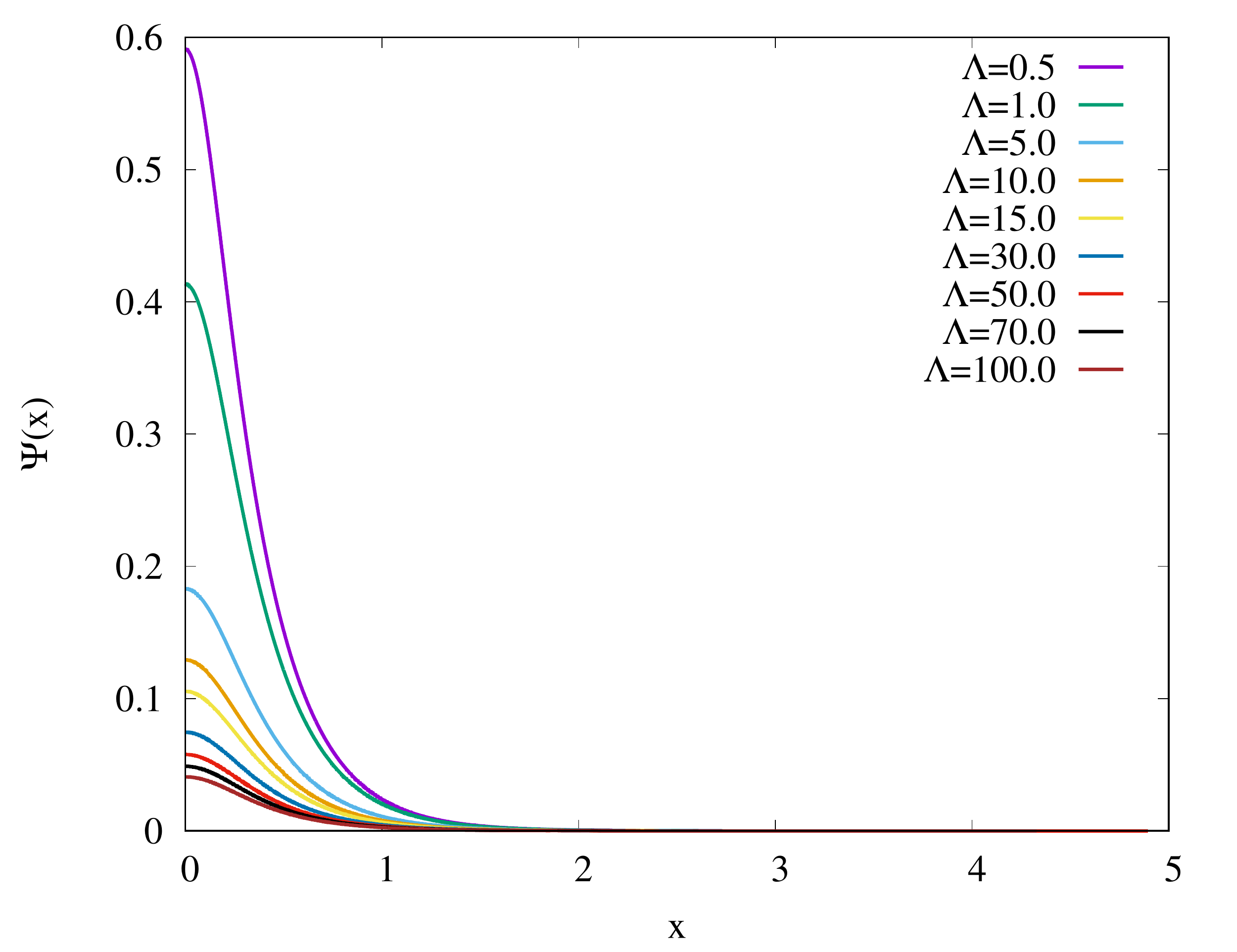}
(d)  \includegraphics[angle =-90,scale=0.3]{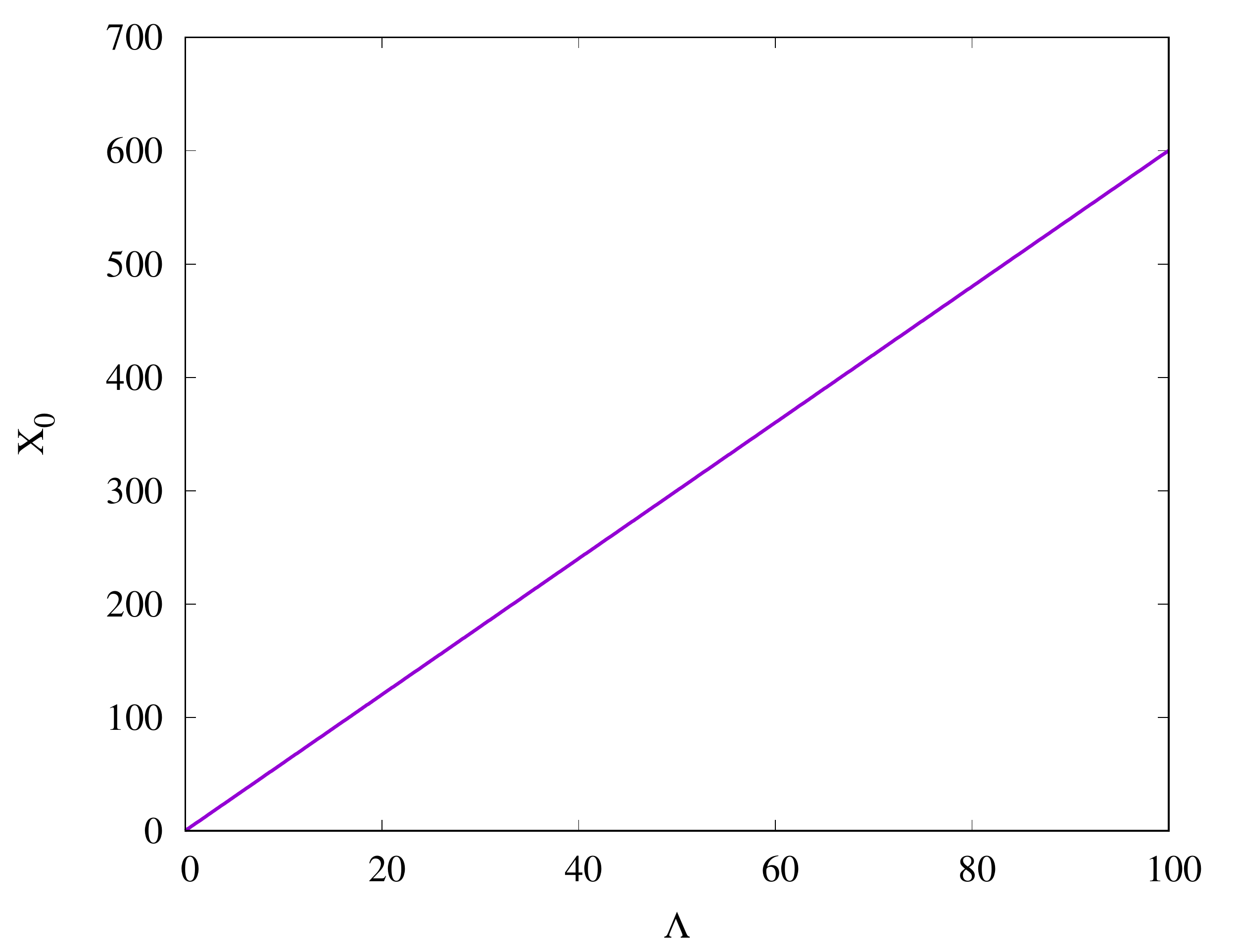}
}
\mbox{
(e) \includegraphics[angle =-90,scale=0.3]{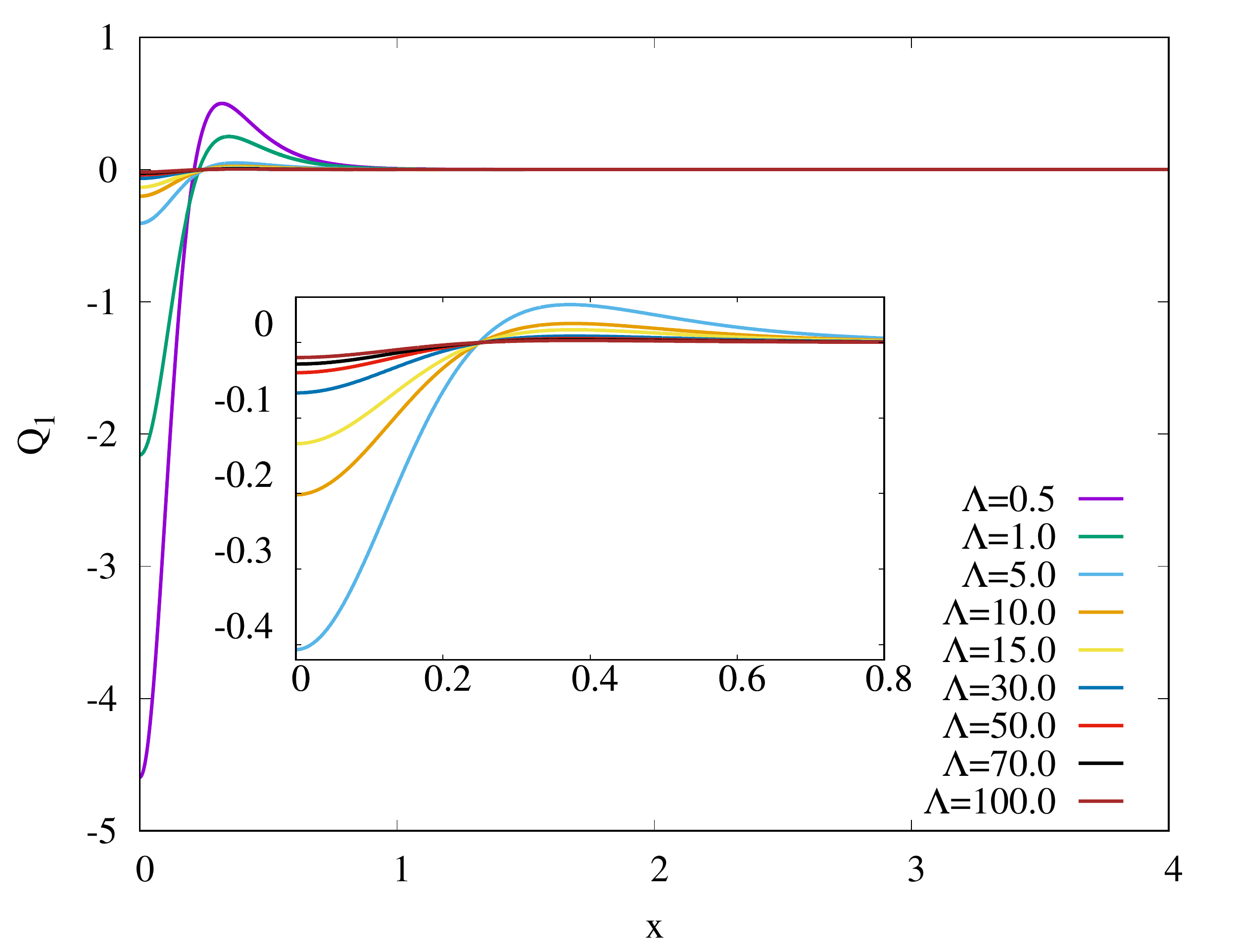}
(f)  \includegraphics[angle =-90,scale=0.3]{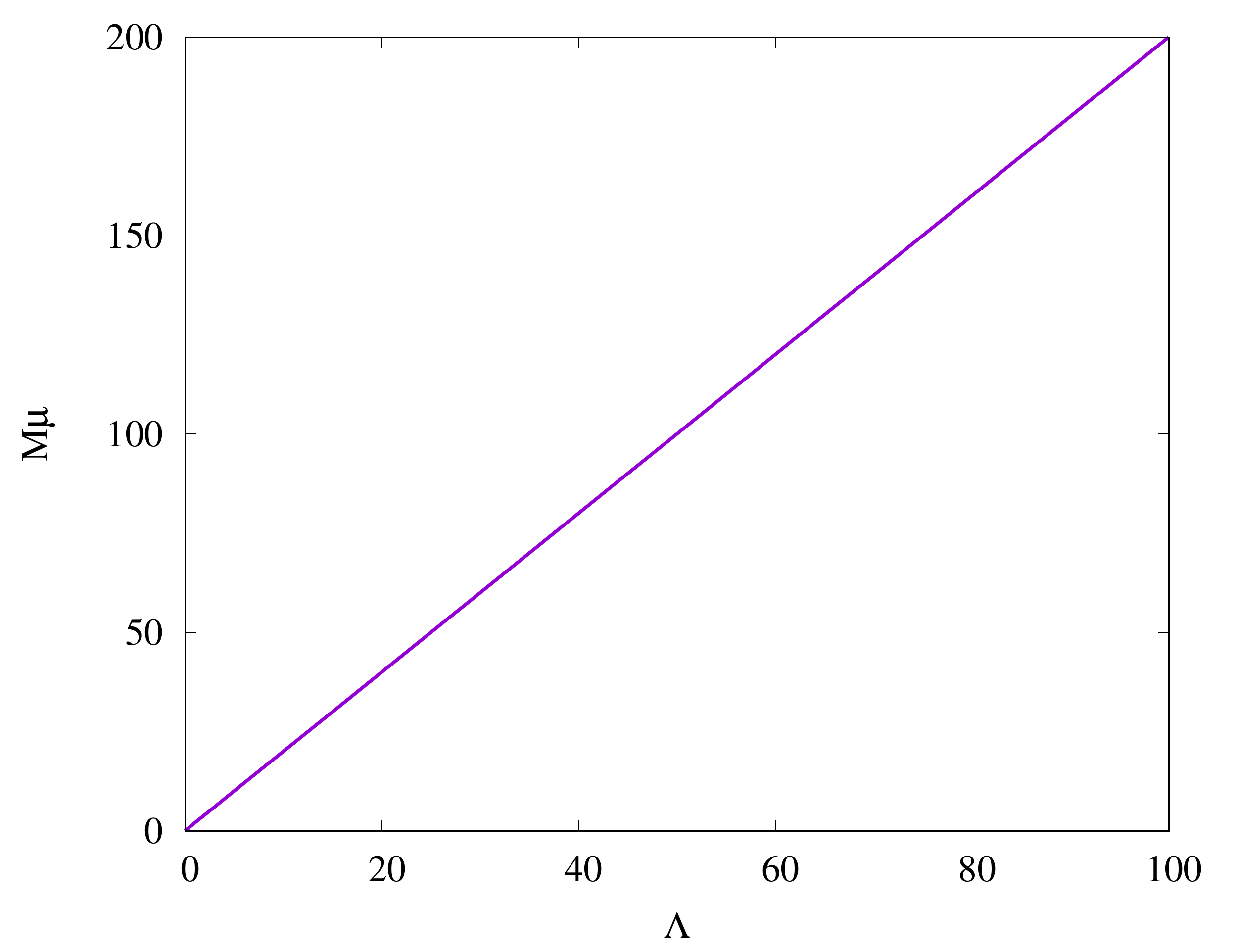}
}
\caption{The function (a) $P$ and (b) $X$ in the logarithmically scaled coordinate of $x$ for several values of $\Lambda$; (c) The function $\Psi$ for several values of $\Lambda$; (d) The circumferential radius of wormhole at $x=0$, i.e., $X_0$, vs $\Lambda$; (e) The violation of the NEC of wormhole. Here, we only show the behavior of $Q_1$, which is given by Eq.~\eqref{Q1Q2exp}, because $Q_2$ is identically zero; (f) The mass $M \mu$ increases almost linearly with $\Lambda$.}
\label{plot_properties}
\end{figure}

\begin{figure}[t!]
\centering
\mbox{(a)\includegraphics[angle =0,scale=0.5]{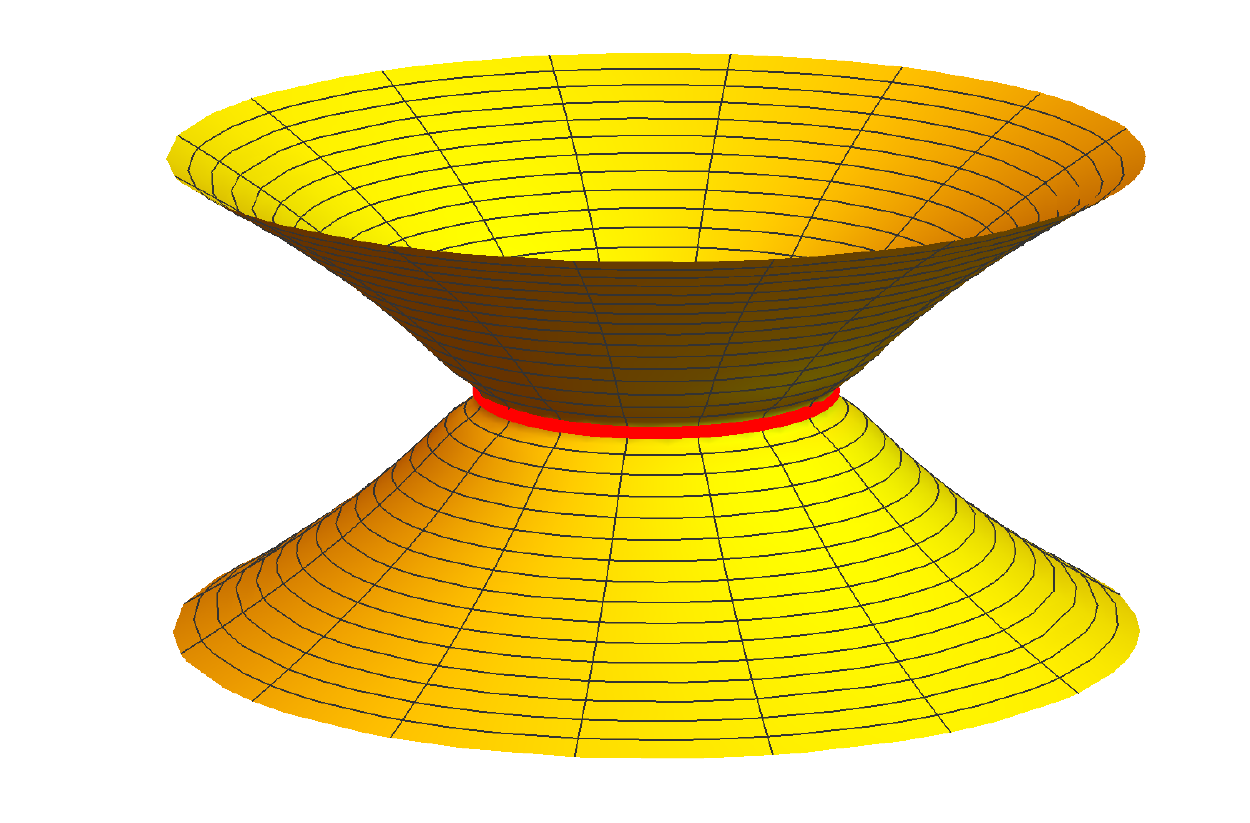}
(b)\includegraphics[angle =0,scale=0.25]{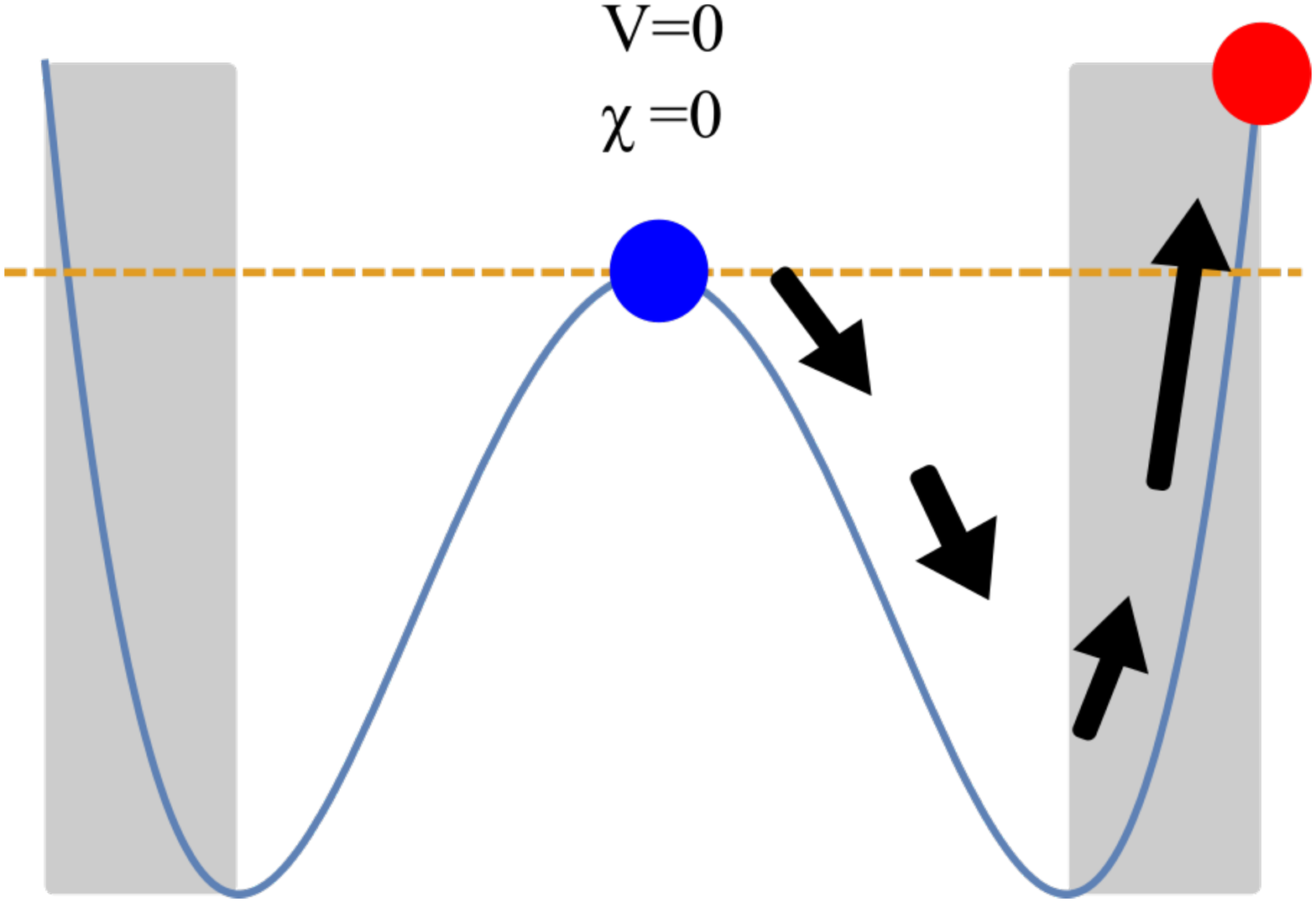}}
\caption{(a) The embedding diagram of the wormhole spacetime supported by the 3-form field. The red curve indicates the wormhole throat. (b) This schematic plot shows the evolution of the 3-form field on the potential in the wormhole spacetime. The shaded regions represent the regions where the NEC is violated.}
\label{fig.embedding}
\end{figure}

\section{Results and Discussion}\label{sec.result}

We exhibit the results of our numerical calculations for $\Lambda$ in the range of $0.01 \leqslant \Lambda \leqslant 100$ in Fig. \ref{plot_properties}. In fact, we find that in order to have a viable wormhole spacetime, $\Lambda$ has to be positive. From Fig. \ref{plot_properties}(c)(d)(f), it can be seen that as $\Lambda$ increases, the mass and the circumferential radius $X_0$ increase almost linearly but the 3-form value at the throat $\Psi_0$ decreases monotonically. Note that $X_0$ is inversely proportional to $\Psi_0$. Furthermore, the value of the redshift function $P$ at $x=0$ does not vary too much for very large values of $\Lambda$ (see Fig.~\ref{plot_properties}(a)). Also, as we have mentioned, the wormhole spacetime contains only a single throat at $x=0$ within the parameter space under consideration. The embedding diagram of the wormhole spacetime constructed here is depicted in Fig.~\ref{fig.embedding}(a). The throat is highlighted with the red curve in this figure.

In GR, the construction of wormholes requires the violation of energy conditions. Here, we will show that the NEC is indeed violated in the wormhole spacetime constructed in this paper, though the $3$-form has the correct sign in its kinetic term. For any null vector $k^\mu$, the NEC requires
\begin{equation}
 T_{\mu \nu} k^\mu k^\nu \geqslant 0 \,.\label{NECcri}
\end{equation}
If there exists a null vector such that the above inequality is not satisfied, then the NEC is violated. Since the wormhole spacetime is spherically symmetric, we consider the following two choices of null (co-)vectors \cite{Antoniou:2019awm}, 
\begin{equation}
k_\mu = \left( g_{tt} , \sqrt{-\frac{g_{tt}}{g_{rr}}}, 0, 0 \right) \,, \quad \text{and} \quad k_\mu = \left( 1, 0, \sqrt{- \frac{g_{tt}}{g_{\theta \theta}}  } ,  0 \right)\,,
\end{equation}
which, following \eqref{NECcri}, yield two different expressions that may be used to measure the violation of NEC:
\begin{equation}
 Q_1:=-T^t\,_t + T^r\,_r  \,, \quad  Q_2:=-T^t\,_t + T^\theta\,_\theta \,.
\end{equation}
In the setup of the wormhole considered here, one can use Eqs.~\eqref{rhoA} and \eqref{pA} to express $Q_1$ and $Q_2$ explicitly
\begin{equation}
  Q_1  = \rho_A+p_{1A} = -\chi\frac{\partial V}{\partial\chi} = 12 \left( \mu^2 - 12 \lambda \chi^2  \right) \chi^2 \,, \quad  Q_2  = \rho_A+p_{2A} = 0 \,.\label{Q1Q2exp}
\end{equation}
Essentially, if one can prove $Q_i$ is negative for a given spacetime, the NEC is then violated. According to Fig.~\ref{plot_properties}(e), the wormhole spacetime considered here indeed violates the NEC. Furthermore, the violation of NEC, i.e., $Q_1<0$, primarily concentrates at the throat, supporting the wormhole spacetime so that it does not collapse. Since the term  $Q_1$ is proportional to the magnitude of $\Psi_0$, which \emph{decreases} monotonically as $\Lambda$ increases, thus the NEC is \emph{less} violated for a large $\Lambda$.

In Fig.~\ref{fig.embedding}(b), we draw a schematic plot to exhibit how the 3-form field, or more precisely, the scalar degree of freedom inherent in the 3-form field, changes on its potential in different regions of the wormhole spacetime. Since $A^2=-6\chi^2$, the 3-form potential \eqref{Va2}, as a function of $\chi$, has a Mexican hat shape. At the asymptotic region, the $\chi$ field vanishes and it is located on the local maximum of the potential $V=0$ (the blue circle). When moving toward the throat, the $\chi$ field grows and it rolls down the potential. After passing through the local minimum, the field climbs up the potential, reaching a finite value where the potential is larger than zero (red circle). This point is the wormhole throat, and the potential value at this point is solely determined by $X_0$: $V|_{\textrm{throat}}=\mu^2/\kappa X_0^2$. In this figure, we also shade the regions where the NEC is violated, in which the wormhole throat resides.

It is worth mentioning that our wormhole solution possesses similar qualitative behavior with the wormhole proposed in Ref.~\cite{Dzhunushaliev:2017syc}, which is supported by a complex phantom scalar field with a very large value of the coefficient of the quartic self-interaction term $\lambda$. In that model, the wormhole solutions are supported by the Mexican hat type potential of the complex phantom scalar field, which contains the time harmonic dependence. The corresponding wormholes contain only a single throat at the radial coordinate $r=0$ and for a very large value of $\lambda$, their circumferential radius $X_0$ and mass increase linearly, while the scalar field at $r=0$ decreases monotonically. This behavior is similar to what we have found for the 3-form field in our model. However, in the model of Ref.~\cite{Dzhunushaliev:2017syc}, when the quartic coefficient $\Lambda$ is very small, the wormholes possess a non-zero mass and $X_0$, which is in contrast to our case. At this point, we can conclude that, at least under the assumption that the wormhole is symmetric and only has a single throat, the existence of the wormhole supported by the 3-form field strongly relies on the existence of the quartic coefficient in the 3-form potential.

\section{Photon sphere, shadow, and ISCO}\label{sec.geo}
In this section, 
we will investigate the geodesic equations of particles moving around the wormhole spacetime obtained in this paper. In particular, we will find the locations of the photon sphere and the ISCO. We will also investigate the apparent size of the shadow cast by this wormhole, and compare the results with those of the Schwarzschild black hole. We would like to mention that the shadows of other wormhole spacetimes have been investigated in various papers \cite{Bambi:2013nla,Nedkova:2013msa,Shaikh:2018kfv,Gyulchev:2018fmd,Bambi:2021qfo,Narzilloev:2021ygl,Jusufi:2021lei}. Other interesting astrophysical observations regarding wormholes, such as quasi-periodic oscillations, have also been investigated \cite{Deligianni:2021ecz,DeFalco:2021btn,Deligianni:2021hwt}.

Because the wormhole spacetime is static and spherically symmetric, the corresponding Killing vectors give two constants of motion:
\begin{equation}
\dot{t}=\frac{E}{P^2}\,,\qquad \dot\varphi=\frac{L}{R^2}\,,
\end{equation}
where $E$ and $L$ are the energy and angular momentum of the particle. The dot denotes the derivative with respect to the proper time of timelike particles, or an affine parameter for massless particles. Due to the spherical symmetry, the particles always undergo planar motion, thus without loss of generality, we can restrict the motion on the equatorial plane ($\theta=\pi/2$). The radial geodesic equation can then be written as
\begin{equation}
\dot{r}^2=\frac{E^2}{P^2}-\frac{L^2}{R^2}+\delta\,,\label{radialgeo}
\end{equation}
where $\delta=-1$ for timelike geodesics and $\delta=0$ for null geodesics, respectively.

Since the metric function $P(r)$ is well-defined and does not vanish for all $r$, we can rewrite the radial geodesic equation \eqref{radialgeo} as follows
\begin{equation}
P^2\dot{r}^2=E^2-V_{\textrm{eff}}\,,
\end{equation}
where the effective potential is defined as
\begin{equation}
V_{\textrm{eff}}:= P^2\left(\frac{L^2}{R^2}-\delta\right)\,.
\end{equation}

We now study the case for null geodesics and timelike geodesics in turn:

\begin{enumerate}
\item \underline{Null geodesics ($\delta=0$) and the photon sphere:}\newline
\newline
When $\delta=0$, the effective potential becomes
\begin{equation}
V_{\textrm{eff}}=\frac{P^2}{R^2}L^2\,.\label{effepophoton}
\end{equation}
It can be shown that the effective potential \eqref{effepophoton} has a local maximum at $r=0$. Therefore, the photon sphere of the wormhole considered in this work is located at the throat of the wormhole.

\subsection*{Remark on Black Hole Shadow}
The existence of the photon sphere enables the wormhole spacetime considered in this paper to cast a shadow. The contour of the shadow is defined by the impact parameter of the photon sphere. The impact parameter $b$ is defined in the usual manner by \cite{Synge:1966okc}
\begin{equation}
b:= \frac{L}{E}\,.
\end{equation}
Since the photon sphere is located at the throat of the wormhole, the impact parameter of the photons which are trapped on the photon sphere, or the apparent radius $R_s$ of the shadow contour, is given by \cite{Synge:1966okc}
\begin{equation}
R_s=\frac{R}{P}\Big|_{r=0}\,.
\end{equation}

For the Schwarzschild black hole, the photon sphere is located at $R=3M$ with $M$ being the black hole mass. The apparent size of its shadow is $R_s=3\sqrt{3}M$. We then calculate the apparent shadow size of the wormhole spacetime obtained in this work and compare the results with that of the Schwarzschild black hole. The mass of the wormhole is given by Eq.~\eqref{whmass}. The results are shown by the cyan curve in Fig.~\ref{fig.isco}(a). It can be seen that when $\Lambda$ is large, the apparent size of the wormhole shadow approaches to that of the Schwarzschild black hole (the brown dotted line). See also Fig.~\ref{fig.isco}(b). Note that Fig.~\ref{fig.isco}(b) also shows that the circumferential radius of the wormhole throat approaches $3M$ when $\Lambda$ is large. On the other hand, as $\Lambda$ decreases, the shadow size shrinks and can be constrained using the shadow images when a more accurate size measurement of the photon light ring is available. This could be achieved in the next generation EHT upgrades in the future. Note that in several other wormhole models, the shadow size of the wormholes has been discovered to be also smaller than the black hole counterparts with the same mass \cite{Gyulchev:2018fmd,Amir:2018pcu,Brahma:2020eos}.

\item \underline{Timelike geodesics ($\delta=-1$) and the ISCO:}\newline\newline
When $\delta=-1$, the effective potential for timelike particles can be written as
\begin{equation}
V_{\textrm{eff}}=P^2\left(\frac{L^2}{R^2}+1\right)\,.\label{effepotimelike}
\end{equation}
After differentiating the effective potential \eqref{effepotimelike} twice, one gets
\begin{align}
V_{\textrm{eff}}'&=2P^2\left(\frac{P'}{P}+\frac{L^2 P'}{R^2 P}-\frac{L^2 R'}{R^3}\right)\,,\\
V_{\textrm{eff}}''&=2P^2\left(\frac{P'^2}{P^2}+\frac{P''}{P}-\frac{L^2R''}{R^3}+\frac{3L^2R'^2}{R^4}+\frac{L^2P''}{R^2 P}+\frac{L^2 P'^2}{R^2 P^2}-\frac{4L^2 P' R'}{R^3 P}\right)\,.
\end{align}
The ISCO is the radius on which $V_{\textrm{eff}}'=V_{\textrm{eff}}''=0$. After some algebraic simplification, the ISCO is located at the radii which are the root of the following differential equation
\begin{equation}
\frac{3P'^2}{P^2}-\frac{3P'R'}{PR}-\frac{P''}{P}+\frac{P'R''}{PR'}=0\,.
\end{equation}
In Fig.~\ref{fig.isco}(a), we vary the value of $\Lambda$ and show the corresponding radius of the ISCO, rescaled by the mass of the wormhole (purple curve). We find that for small values of $\Lambda$, the ISCO radius, $R_{\textrm{ISCO}}/M$, is larger than that of the Schwarzschild black hole. As $\Lambda$ increases, the Schwarzschild result, i.e., $R_{\textrm{ISCO}}/M=6$ (the green dotted line), is recovered. See also the purple line in Fig.~\ref{fig.isco}(b).

\begin{figure}[t!]
\centering

\centering
\mbox{(a)\includegraphics[angle =-90,scale=0.3]{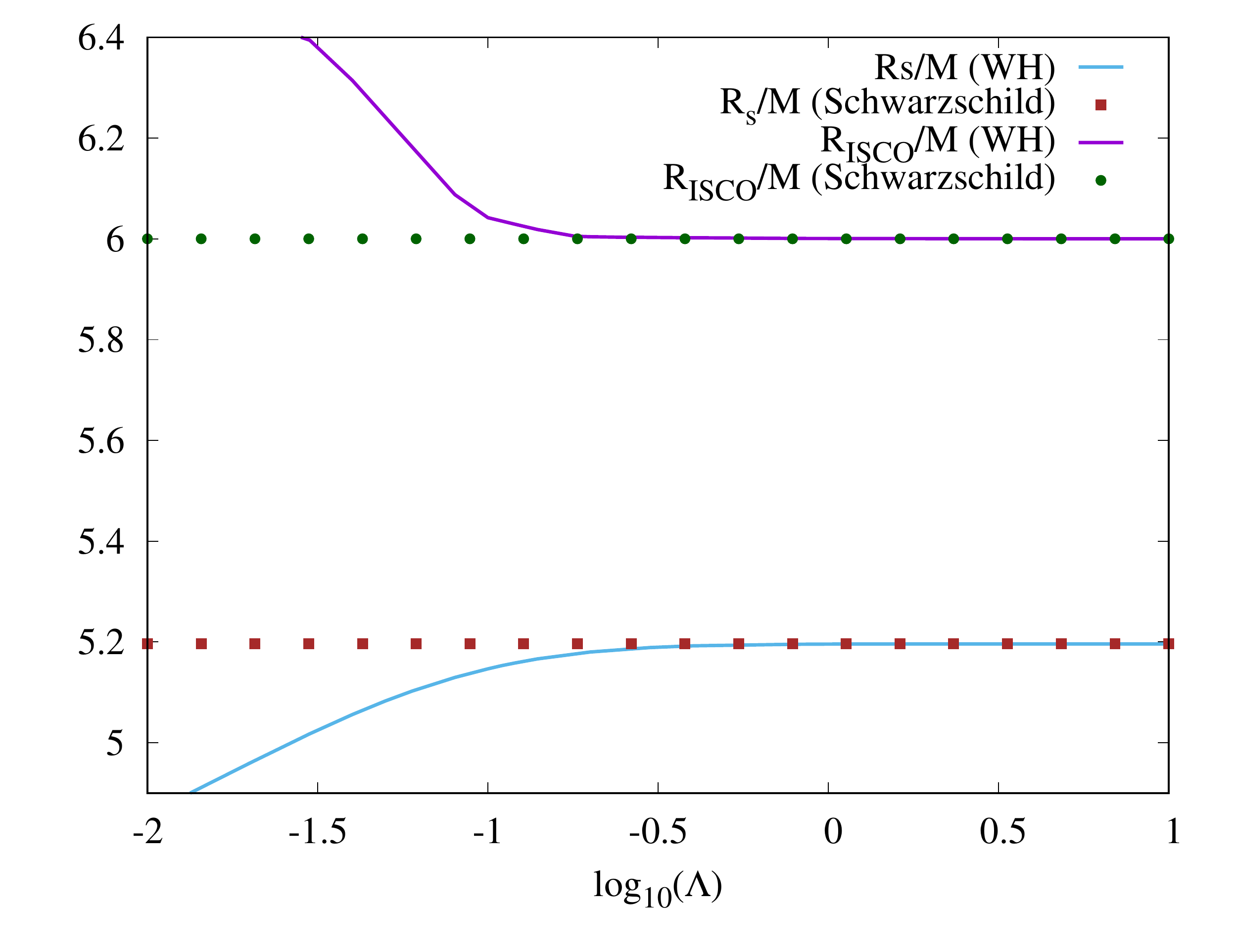}
(b)\includegraphics[angle =-90,scale=0.3]{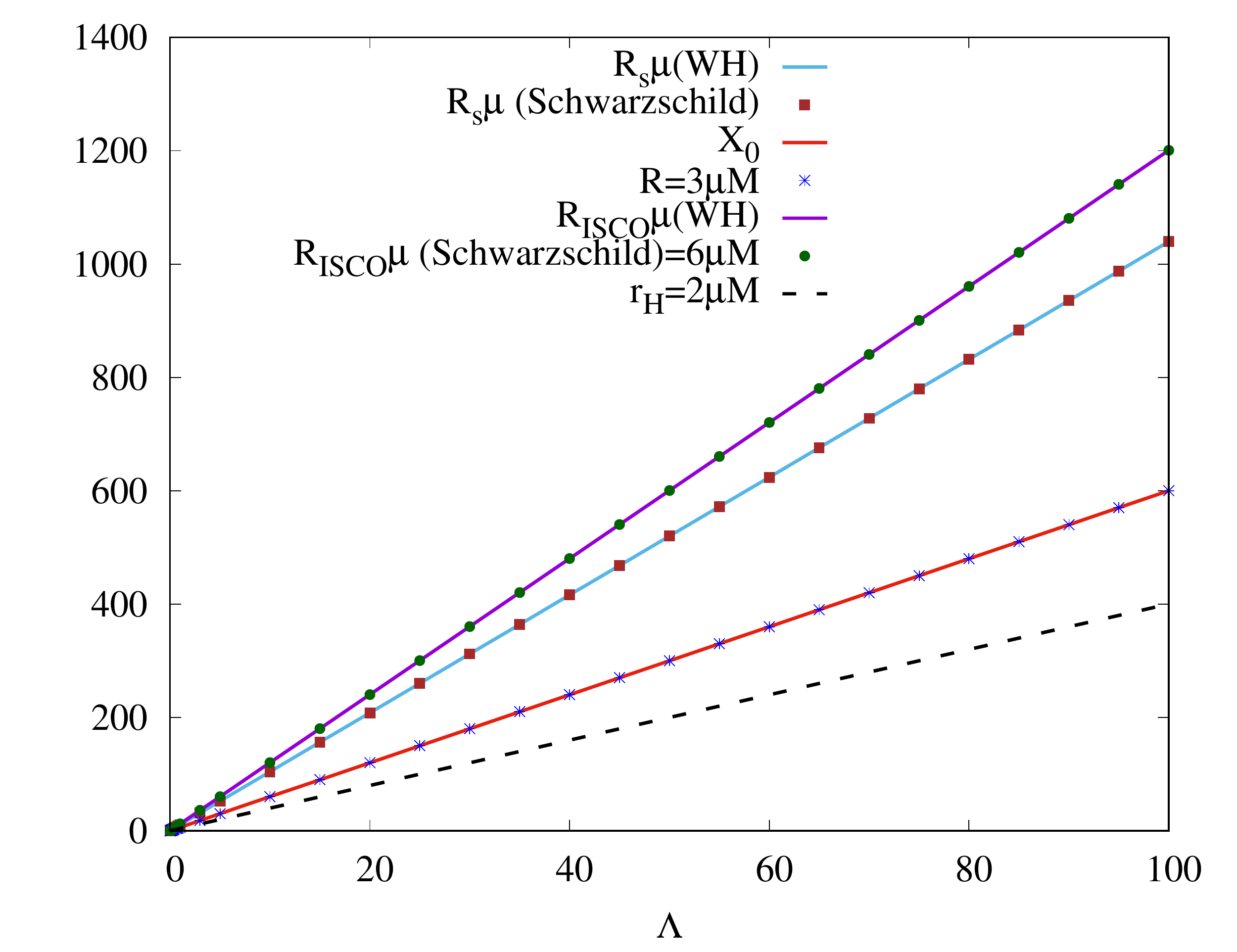}}
\caption{(a) The apparent radius $R_s/M$ of the wormhole shadow (cyan) and the radius of ISCO $R_{\textrm{ISCO}}/M$ (purple) are shown with respect to $\Lambda$. Note that $M$ denotes the mass of the wormhole. (b) When $\Lambda$ is large, the apparent shadow size, the radius of the photon sphere, and the ISCO radius of the wormhole all reduce to those of the Schwarzschild results, i.e., $R_s\rightarrow 3\sqrt{3}M$, $R_0\rightarrow 3M$, $R_{\textrm{ISCO}}\rightarrow6M$.}
\label{fig.isco}
\end{figure}

\end{enumerate}

Our results indicate that, as $\Lambda$ is large enough, the location of the ISCO and apparent size of the shadow reduce to those of the Schwarzschild black hole. Therefore, the astrophysical observations which rely on testing the geodesic motions of the particles moving around the wormhole will be very challenging to distinguish the $3$-form wormhole from a Schwarzschild black hole, meaning that the 3-form wormhole constructed in this paper could be a potential black hole mimicker. This does not rule out the possibility of distinguishing such wormholes from black holes if we consider more subtle influences from gravitational perturbations from the other side of the wormhole \cite{1910.00429, 2007.12184}.

\section{Conclusion and Outlook}\label{sec:con}

We have constructed a symmetric wormhole solution in GR, which is supported by a 3-form field with a potential that contains a quartic self-interaction term. The wormhole solution only contains a single throat, which is located at the radial coordinate $r=0$. Although the NEC is violated near the throat, we emphasize that the kinetic term of the 3-form has the right sign. As the coefficient of the quartic self-interaction term, i.e., $\Lambda$, increases, the wormhole mass and its circumferential radius at the throat increase almost linearly, while the 3-form field value at the throat decreases. For very large values of $\Lambda$, these wormholes behave qualitatively similar to other wormholes supported by different matter fields that possess quartic self-interaction term in the potential. This includes, for example, a complex phantom field with the harmonic time dependence \cite{Dzhunushaliev:2017syc}. However, the latter behaves differently for very small values of the coefficient of quartic self-interaction term, in the sense that the corresponding wormhole solution with a complex phantom field possesses finite values of the mass and circumferential radius at the throat. In our case, however, a non-zero $\Lambda$ is necessary in order to support the wormhole throat. This can also be seen from the expression $Q_1$ in Eq.~\eqref{Q1Q2exp} that the energy condition is violated only when $\Lambda>0$.

We also investigate the geodesic equations for null particles and timelike particles moving around the wormhole. It is found that the unstable photon sphere, on which photons can undergo circular motions around the wormhole, is exactly at the wormhole throat. In addition, the wormhole is able to cast a shadow contour nearly indistinguishable from that of the Schwarzschild black hole, as long as $\Lambda$ is large enough. We also investigate the locations of the ISCO for massive particles, and it is found that the radius of ISCO deviates from the Schwarzschild counterpart when $\Lambda$ is small, but reduces to it for a larger $\Lambda$. Thus, our wormholes can be a black hole mimicker when $\Lambda$ is large, precisely when NEC is less violated. Of course, most astrophysical black holes rotate, so it remains to be seen if this mimicry still holds when rotation is considered.  

We remark that the stability of the 3-form wormhole presented in this work is still yet to be explored. Future investigation should look into the radial perturbation on the background metric and the form field to check stability, among other considerations. Indeed, the wormhole solutions supported by the complex phantom scalar field are found to be unstable against linear perturbations \cite{Dzhunushaliev:2017syc}. It would be interesting to check whether our wormholes suffer from the same instability, and if so, for which range of $\Lambda$.


\acknowledgments

The work of M.B.L. is supported by the Basque Foundation of Science Ikerbasque. She
also would like to acknowledge the partial support from the Basque government Grant No. IT956-16 (Spain) and
from the project FIS2017-85076-P (MINECO/AEI/FEDER, UE). Her work has been supported by the Spanish project PID2020-114035GB-100  (MINECO/AEI/FEDER, UE). She is as grateful to the kind invitation of the Center for Gravitation and Cosmology of Yangzhou University where this project was initiated back in December 2019. C.Y.C is supported by the Institute of Physics of Academia Sinica. D.Y and X.Y.C are supported by the National Research Foundation of Korea (Grant No.:  2021R1C1C1008622, 2021R1A4A5031460). Y.C.O. thanks NNSFC (grant No.11922508) for
funding support. XYC acknowledges the discussion with Jutta Kunz.



\begin{thebibliography}{99}

\bibitem{Einstein:1935tc}
  A.~Einstein, N.~Rosen,
  ``The Particle Problem in the General Theory of Relativity'', {Phys.\ Rev.\  {\bf 48} (1935) 73}. 


\bibitem{Visser:1995cc} 
  M.~Visser,
  \emph{Lorentzian Wormholes: From Einstein to Hawking}, Woodbury, USA: AIP, 1995. 

\bibitem{visser1}
C.~Barcelo, M.~Visser, ``Traversable Wormholes From Massless Conformally Coupled Scalar Fields'', {Phys. Lett. B \textbf{466} (1999) 127}. 

\bibitem{visser2}
M.~Visser, ``Traversable Wormholes: Some Simple Examples'', Phys. Rev. D \textbf{39} (1989) 3182(R) 

\bibitem{MTY}
M.~S.~Morris, K.~S.~Thorne, U.~Yurtsever, ``Wormholes, Time Machines, and the Weak Energy Condition'', {Phys. Rev. Lett. \textbf{61} (1988) 1446} 

\bibitem{1405.0403}
E.~Curiel, ``A Primer on Energy Conditions'', {Einstein Stud. \textbf{13} (2017) 43}, in \emph{Towards a Theory of Spacetime Theories}, eds. D. Lehmkuhl, G. Schiemann, and E. Scholz, Einstein Studies, Birkh\"auser, 2017, ch. 3. 


\bibitem{Ellis:1973yv}
  H.~G.~Ellis,
  ``Ether Flow Through a Drainhole -- A Particle Model In General Relativity'', {J.\ Math.\ Phys.\  {\bf 14} (1973) 104}

\bibitem{Ellis:1979bh}
  H.~G.~Ellis,
  ``The Evolving, Flowless Drain Hole: A Nongravitating Particle Model In General Relativity Theory'', {Gen.\ Rel.\ Grav.\  {\bf 10} (1979) 105}.

\bibitem{Bronnikov:1973fh}
  K.~A.~Bronnikov,
  ``Scalar-Tensor Theory and Scalar Charge'',
  Acta Phys.\ Polon.\  {\bf B4} (1973) 251.





\bibitem{0302506}
R.~R.~Caldwell, M.~Kamionkowski, N.~N.~Weinberg, ``Phantom Energy and Cosmic Doomsday'', Phys. Rev. Lett. \textbf{91} (2003) 071301. 





\bibitem{Harko:2013yb}
T.~Harko, F.~S.~N.~Lobo, M.~K.~Mak and S.~V.~Sushkov,
``Modified-gravity wormholes without exotic matter,''  Phys. \ Rev. \ D \textbf{87}, no.6, 067504 (2013).


\bibitem{Kanti:2011yv}
  P.~Kanti, B.~Kleihaus, J.~Kunz,
  ``Stable Lorentzian Wormholes in Dilatonic Einstein-Gauss-Bonnet Theory'', Phys.\ Rev.\ D {\bf 85} (2012) 044007.

\bibitem{Antoniou:2019awm}
G.~Antoniou, A.~Bakopoulos, P.~Kanti, B.~Kleihaus, J.~Kunz,
``Novel Einstein–Scalar-Gauss-Bonnet Wormholes Without Exotic Matter'', Phys. Rev. D \textbf{101} (2020) no.2, 024033.



\bibitem{Brihaye:2020dgo}
Y.~Brihaye, J.~Renaux,
``Scalarized-Charged Wormholes in Einstein-Gauss-Bonnet Gravity'', [arXiv:2004.12138 [gr-qc]]


\bibitem{Ibadov:2020btp}
R.~Ibadov, B.~Kleihaus, J.~Kunz, S.~Murodov,
``Wormholes in Einstein-Scalar-Gauss-Bonnet Theories With a Scalar Self-Interaction Potential'', Phys. Rev. D \textbf{102} (2020) no.6, 064010.


\bibitem{Chakraborty:2021qwl}
K.~Chakraborty, A.~Aziz, F.~Rahaman and S.~Ray,
[arXiv:2104.13966 [gr-qc]].



\bibitem{Mironov:2018uou}
S.~Mironov, V.~Rubakov, V.~Volkova,
``More About Stable Wormholes in Beyond Horndeski Theory'', Class. Quant. Grav. \textbf{36} (2019) no.13, 135008.




\bibitem{Volkova:2019kyd}
V.~Volkova,
``Searching for Wormholes Beyond Horndeski Theories'', Universe \textbf{5} (2019) no.2, 54. 



\bibitem{Korolev:2020ohi}
R.~Korolev, F.~S.~Lobo, S.~V.~Sushkov,
``General Constraints on Horndeski Wormhole Throats'', Phys. Rev. D \textbf{101} (2020) 124057.


\bibitem{Li:2020jyf}
A.~C.~Li and X.~F.~Li,
``Morris-Thorne wormhole in the vector-tensor theories with Abelian gauge symmetry breaking,'' Phys. Rev. D \textbf{104}, no.4, 044006 (2021).



\bibitem{Rosa:2018jwp}
J.~L.~Rosa, J.~P.~S.~Lemos and F.~S.~N.~Lobo,
``Wormholes in generalized hybrid metric-Palatini gravity obeying the matter null energy condition everywhere,''
Phys. Rev. D \textbf{98} (2018) no.6, 064054



\bibitem{Rosa:2021yym}
J.~L.~Rosa,
``Double gravitational layer traversable wormholes in hybrid metric-Palatini gravity,''
[arXiv:2107.14225 [gr-qc]].



\bibitem{Battarra:2014naa}
L.~Battarra, G.~Lavrelashvili and J.~L.~Lehners,
``Creation of wormholes by quantum tunnelling in modified gravity theories'', Phys. Rev. D \textbf{90}, no.12, 124015 (2014).




\bibitem{Bouhmadi-Lopez:2018sto}
M.~Bouhmadi-L\'opez, C.~Y.~Chen, P.~Chen and D.~h.~Yeom,
``Regular Instantons in the Eddington-inspired-Born-Infeld Gravity: Lorentzian Wormholes from Bubble Nucleations'', JCAP \textbf{10}, 056 (2018).



\bibitem{Tumurtushaa:2018agq}
G.~Tumurtushaa and D.~H.~Yeom,
``Quantum creation of traversable wormholes ex nihilo in Gauss\textendash{}Bonnet-dilaton gravity'',
{Eur. Phys. J. C \textbf{79}, no.6, 488 (2019)}.

\bibitem{Chew:2020lkj}
X.~Y.~Chew, G.~Tumurtushaa and D.~h.~Yeom,
``Euclidean wormholes in Gauss\textendash{}Bonnet-dilaton gravity,''
Phys. Dark Univ. \textbf{32} (2021), 100811



\bibitem{Blazquez-Salcedo:2019uqq}
J.~L.~Bl\'azquez-Salcedo, C.~Knoll,
``Constructing spherically symmetric Einstein\textendash{}Dirac systems with multiple spinors: Ansatz, wormholes and other analytical solutions'', Eur. Phys. J. C \textbf{80}, no.2 (2020) 174.




\bibitem{Blazquez-Salcedo:2020czn}
J.~L.~Bl\'azquez-Salcedo, C.~Knoll, E.~Radu,
``Traversable Wormholes in Einstein-Dirac-Maxwell Theory'', Phys. Rev. Lett. \textbf{126} (2021) no.10, 101102.



\bibitem{Konoplya:2021hsm}
R.~A.~Konoplya and A.~Zhidenko,
``Traversable wormholes in General Relativity without exotic matter,''
[arXiv:2106.05034 [gr-qc]].


\bibitem{Copeland:1994km}
E.~J.~Copeland, A.~Lahiri and D.~Wands,
``String cosmology with a time dependent antisymmetric tensor potential,''
Phys. Rev. D \textbf{51}, 1569-1576 (1995).


\bibitem{Lukas:1996iq}
A.~Lukas, B.~A.~Ovrut and D.~Waldram,
``String and M theory cosmological solutions with Ramond forms,''
Nucl. Phys. B \textbf{495}, 365-399 (1997).

\bibitem{Ovrut:1997ur}
B.~A.~Ovrut and D.~Waldram,
``Membranes and three form supergravity,''
Nucl. Phys. B \textbf{506}, 236-266 (1997).

\bibitem{Farakos:2017ocw}
F.~Farakos, S.~Lanza, L.~Martucci and D.~Sorokin,
``Three-forms, Supersymmetry and String Compactifications,''
Phys. Part. Nucl. \textbf{49}, no.5, 823-828 (2018).


\bibitem{Duff:1980qv}
M.~J.~Duff and P.~van Nieuwenhuizen,
``Quantum Inequivalence of Different Field Representations,''
Phys. Lett. B \textbf{94}, 179-182 (1980).


\bibitem{Turok:1998he}
N.~Turok and S.~W.~Hawking,
``Open inflation, the four form and the cosmological constant,''
Phys. Lett. B \textbf{432}, 271-278 (1998).


\bibitem{Germani:2009iq}
C.~Germani and A.~Kehagias,
``P-nflation: generating cosmic Inflation with p-forms,''
JCAP \textbf{03}, 028 (2009).

\bibitem{Koivisto:2009sd}
T.~S.~Koivisto, D.~F.~Mota and C.~Pitrou,
``Inflation from N-Forms and its stability,''
JHEP \textbf{09}, 092 (2009).

\bibitem{Germani:2009gg}
C.~Germani and A.~Kehagias,
``Scalar perturbations in p-nflation: the 3-form case,''
JCAP \textbf{11}, 005 (2009).

\bibitem{Koivisto:2009fb}
T.~S.~Koivisto and N.~J.~Nunes,
``Inflation and dark energy from three-forms,''
Phys. Rev. D \textbf{80}, 103509 (2009).

\bibitem{Koivisto:2009ew}
T.~S.~Koivisto and N.~J.~Nunes,
``Three-form cosmology,''
Phys. Lett. B \textbf{685}, 105-109 (2010).

\bibitem{DeFelice:2012jt}
A.~De Felice, K.~Karwan and P.~Wongjun,
``Stability of the 3-form field during inflation,''
Phys. Rev. D \textbf{85}, 123545 (2012).

\bibitem{DeFelice:2012wy}
A.~De Felice, K.~Karwan and P.~Wongjun,
``Reheating in 3-form inflation,''
Phys. Rev. D \textbf{86}, 103526 (2012).


\bibitem{Kumar:2014oka}
K.~S.~Kumar, J.~Marto, N.~J.~Nunes and P.~V.~Moniz,
``Inflation in a two 3-form fields scenario,''
JCAP \textbf{06}, 064 (2014).

\bibitem{Mulryne:2012ax}
David J. Mulryne, Johannes Noller, Nelson J. Nunes, ``Three-Form Inflation and Non-Gaussianity'', JCAP \textbf{1212} (2012) 016.	

\bibitem{SravanKumar:2016biw}
K.~Sravan Kumar, D.~J.~Mulryne, N.~J.~Nunes, J.~Marto and P.~Vargas Moniz,
``Non-Gaussianity in multiple three-form field inflation,''
Phys. Rev. D \textbf{94}, no.10, 103504 (2016).







\bibitem{Morais:2016bev}
J.~Morais, M.~Bouhmadi-L\'opez, K.~Sravan Kumar, J.~Marto, Y.~Tavakoli,
``Interacting 3-Form Dark Energy Models: Distinguishing Interactions and Avoiding the Little Sibling of the Big RIp'', Phys. Dark Univ. \textbf{15} (2017) 7.

\bibitem{Bouhmadi-Lopez:2016dzw}
M.~Bouhmadi-L\'opez, J.~Marto, J.~Morais and C.~M.~Silva,
``Cosmic infinity: A dynamical system approach,''
JCAP \textbf{03}, 042 (2017).

\bibitem{Bouhmadi-Lopez:2018lly}
M.~Bouhmadi-L\'opez, D.~Brizuela and I.~Garay,
``Quantum behavior of the ''Little Sibling'' of the Big Rip induced by a three-form field,''
JCAP \textbf{09}, 031 (2018).


\bibitem{Bouhmadi-Lopez:2020wve}
M.~Bouhmadi-L\'opez, C.~Y.~Chen, X.~Y.~Chew, Y.~C.~Ong and D.~H.~Yeom,
``Regular Black Hole Interior Spacetime Supported by Three-Form Field,''
Eur. Phys. J. C \textbf{81}, no.4, 278 (2021).

\bibitem{CANTATA:2021ktz}
E.~N.~Saridakis \textit{et al.} [CANTATA],
``Modified Gravity and Cosmology: An Update by the CANTATA Network,''
[arXiv:2105.12582 [gr-qc]].




\bibitem{Barros:2018lca} 
  B.~J.~Barros, F.~S.~N.~Lobo,
  ``Wormhole Geometries Supported By Three-Form Fields'', Phys.\ Rev.\ D {\bf 98}, no. 4 (2018) 044012.

\bibitem{Barros:2020ghz}
B.~J.~Barros, B.~Danila, T.~Harko, F.~S.~N.~Lobo, 
``Black Hole and Naked Singularity Geometries Supported by Three-Form Fields'', Eur. Phys. J. C \textbf{80} (2020) 617.


\bibitem{Barros:2021jbt}
B.~J.~Barros, Z.~Haghani, T.~Harko and F.~S.~N.~Lobo,
``Static spherically symmetric three-form stars,''
Eur. Phys. J. C \textbf{81}, no.4, 307 (2021).


\bibitem{Dzhunushaliev:2017syc} 
  V.~Dzhunushaliev, V.~Folomeev, B.~Kleihaus, J.~Kunz,
  ``Wormhole Solutions With a Complex Ghost Scalar Field and Their Instability'', Phys.\ Rev.\ D {\bf 97}, no. 2 (2018) 024002.
  
 
  
 
\bibitem{Misner:1964je}
C.~W.~Misner,  D.~H.~Sharp,
``Relativistic Equations for Adiabatic, Spherically Symmetric Gravitational Collapse'', Phys. Rev. \textbf{136} (1964) B571. 
 
 

\bibitem{Bambi:2013nla}
C.~Bambi,
``Can the supermassive objects at the centers of galaxies be traversable wormholes? The first test of strong gravity for mm/sub-mm very long baseline interferometry facilities,''
Phys. Rev. D \textbf{87}, 107501 (2013).
 
\bibitem{Nedkova:2013msa}
P.~G.~Nedkova, V.~K.~Tinchev and S.~S.~Yazadjiev,
``Shadow of a rotating traversable wormhole,''
Phys. Rev. D \textbf{88}, no.12, 124019 (2013).
 
\bibitem{Shaikh:2018kfv}
R.~Shaikh,
``Shadows of rotating wormholes,''
Phys. Rev. D \textbf{98}, no.2, 024044 (2018).
 
\bibitem{Gyulchev:2018fmd}
G.~Gyulchev, P.~Nedkova, V.~Tinchev, S.~Yazadjiev,
``On the Shadow of Rotating Traversable Wormholes'', Eur. Phys. J. C \textbf{78}, no.7 (2018) 544.

 
\bibitem{Bambi:2021qfo}
C.~Bambi and D.~Stojkovic,
``Astrophysical Wormholes,''
Universe \textbf{7} (2021) no.5, 136.
 
\bibitem{Narzilloev:2021ygl}
B.~Narzilloev, D.~Malafarina, A.~Abdujabbarov, B.~Ahmedov and C.~Bambi,
``Particle motion around a static axially symmetric wormhole,''
[arXiv:2105.09174 [gr-qc]].
 
\bibitem{Jusufi:2021lei}
K.~Jusufi, S.~K., M.~Azreg-A\"\i{}nou, M.~Jamil, Q.~Wu and C.~Bambi,
``Constraining Wormhole Geometries using the Orbit of S2 Star and the Event Horizon Telescope,''
[arXiv:2106.08070 [gr-qc]].
  
  
\bibitem{Deligianni:2021ecz}
E.~Deligianni, J.~Kunz, P.~Nedkova, S.~Yazadjiev and R.~Zheleva,
``Quasiperiodic oscillations around rotating traversable wormholes,''
Phys. Rev. D \textbf{104}, no.2, 024048 (2021).
  
\bibitem{DeFalco:2021btn}
V.~De Falco, M.~De Laurentis and S.~Capozziello,
``Epicyclic frequencies in static and spherically symmetric wormhole geometries,''
Phys. Rev. D \textbf{104}, no.2, 024053 (2021).
  
\bibitem{Deligianni:2021hwt}
E.~Deligianni, B.~Kleihaus, J.~Kunz, P.~Nedkova and S.~Yazadjiev,
``Quasi-periodic Oscillations in Rotating Ellis Wormhole Spacetimes,''
[arXiv:2107.01421 [gr-qc]].
  
  
  
\bibitem{Synge:1966okc}
J.~L.~Synge,
``The Escape of Photons from Gravitationally Intense Stars'', Mon. Not. Roy. Astron. Soc. \textbf{131}, no.3 (1966) 463.

 
 
\bibitem{Amir:2018pcu}
M.~Amir, K.~Jusufi, A.~Banerjee and S.~Hansraj,
``Shadow Images of Kerr-like Wormholes'', Class. Quant. Grav. \textbf{36}, no.21 (2019) 215007.
 
\bibitem{Brahma:2020eos}
S.~Brahma, C.~Y.~Chen and D.~h.~Yeom,
``Testing Loop Quantum Gravity from Observational Consequences of Nonsingular Rotating Black Holes,''
Phys. Rev. Lett. \textbf{126} (2021) no.18, 181301.
 

\bibitem{1910.00429}
D.-C.~Dai, D.~Stojkovic, ``Observing a Wormhole'', Phys. Rev. D \textbf{100} (2019) 8, 083513. 

\bibitem{2007.12184}
J.~H.~Simonetti, M.~J.~Kavic, D.~Minic, D.~Stojkovic, D.-C.~Dai, ``A Sensitive Search for Wormholes'', [arXiv:2007.12184 [gr-qc]] 





\end{thebibliography}
\end{document}